\documentclass[twocolumn,prb,aps,showpacs,amsmath,amssymb,floatfix]{revtex4-1}

\usepackage{url}
\usepackage{amsmath}
\usepackage{amssymb}
\usepackage{graphicx}
\usepackage{hyperref}
\usepackage{dcolumn}
\usepackage{bm}
\usepackage{xcolor}
\usepackage{siunitx}

\begin{document}
\title{First-principles calculations of spin-orbit torques in Mn$_2$Au/heavy-metal bilayers}

\author{Wuzhang Fang}

\author{K. D. Belashchenko}
\affiliation{Department of Physics and Astronomy and Nebraska Center for Materials and Nanoscience, University of Nebraska-Lincoln, Lincoln, Nebraska 68588, USA}

\begin{abstract}
Using the non-equilibrium Green's function technique, we calculate spin-orbit torques in a Mn$_2$Au/heavy-metal bilayer, where the heavy metal (HM) is W or Pt.
Spin-orbit coupling (SOC) in the bulk of Mn$_2$Au generates a strong fieldlike torquance, which is parallel on the two sublattices and scales linearly with the conductivity, and a weaker
dampinglike torquance that is antiparallel on the two sublattices. Interfaces with W or Pt generate parallel dampinglike torques of opposite signs that are similar in magnitude to those in ferromagnetic bilayers and similarly insensitive to disorder. The dampinglike torque efficiency depends strongly on the termination of the interface and on the presence of spin-orbit coupling in Mn$_2$Au, suggesting that the dampinglike torque is not due solely to the spin-Hall effect in the HM layer. Interfaces also induce antiparallel fieldlike and dampinglike torques that can penetrate deep into Mn$_2$Au.
\end{abstract}

\maketitle

\section{Introduction}

Antiferromagnets (AFM) are promising materials for spintronic applications due to their insensitivity to external magnetic fields and ultrafast spin dynamics. \cite{Baltz-RMP,AFM2,Kampfrath}
Antiferromagnetic order and spin texture, such as domain walls and skyrmions, can be manipulated by current-induced spin-orbit torques (SOT). \cite{ManchonRMP}
Switching of AFM order by fieldlike (FL) SOT was demonstrated experimentally in metallic CuMnAs \cite{CuMnAs} and Mn$_2$Au. \cite{Mn2Au2} These tetragonal compounds have high N\'eel temperatures and collinear AFM structures that have an inversion center only in combination with time reversal. \cite{Mn2Au1,CuMnAs} Such magnetic symmetry allows finite current-induced spin accumulations on the two magnetic sublattices. \cite{Zelezny1,Zelezny2} The staggered component of this spin accumulation exerts a FL SOT that acts in the same direction on the two sublattices. Such torque is enhanced by exchange coupling and can efficiently switch the AFM order parameter. \cite{Zelezny2,Jungwirth2016,GomonayNatPhys2018} The nonstaggered component of the spin accumulation is odd in the AFM order parameter and exerts a dampinglike (DL) torque acting in opposite directions on the two sublattices, which is not exchange-enhanced. To avoid confusion with the staggered effective field, we will refer to a torque as parallel (P) or antiparallel (AP) if it has the same or opposite sign on the two sublattices, respectively.

The AFM order can also be manipulated by SOT in bilayers with heavy metals (HM), \cite{Reichlova2015,Manchon2017} which is always allowed regardless of the bulk symmetry. Different mechanisms \cite{GomonayNatPhys2018} were proposed either in the macrospin model with switching driven by PDL \cite{Zarzuela2017,Chen2018} or APDL \cite{Moriyama2018,Fei2021} torques or through the action of both DL and FL torques on domain walls. \cite{Shiino2016,Baldrati2019} A lot of research has focused on AFM/HM bilayers with insulating AFM, \cite{Chen2018,Moriyama2018,Luqiao2019,Baldrati2019,Schreiber2020,BaldratiCoO2020,Meer2021} but SOT-driven switching remains controversial, because the detection of AFM switching through magnetoresistance measurements is subject to artifacts of non-magnetic origin, \cite{Chiang2019,Churikova2020,Matalla-Wagner} while direct observations of magnetic switching can be explained by the thermomagnetoelastic mechanism that does not involve SOT. \cite{Luqiao2019,Meer2021} Current-induced switching was also reported for a Mn$_2$Au(103)/Pt bilayer \cite{Song2019} where the final state was different compared to a single layer of Mn$_2$Au, and in other metallic AFM/HM bilayers. \cite{Dunz2020,Khalili2020,DuttaGupta2020}

Analysis based on spin-diffusion theory \cite{Manchon2017,Baltz-RMP} shows that transverse spin current can diffuse into a collinear two-sublattice metallic AFM. For an AFM/HM bilayer, this theory predicts a combination of APFL and PDL SOT, assuming a magnetically compensated interface.

In this paper, we study SOT in disordered Mn$_2$Au/HM bilayers using the first-principles non-equilibrium Green's function (NEGF) approach. \cite{datta1997book,Nikolic2018} We find that PFL SOT dominates in the bulk of Mn$_2$Au, but there is also an appreciable APDL component. Interfaces with Pt and W generate all types of SOT, including PDL which is localized at the interface and APFL which can penetrate deep into Mn$_2$Au. 

\section{Computational methods}

\begin{figure}[htb]
\includegraphics[width=0.9\columnwidth]{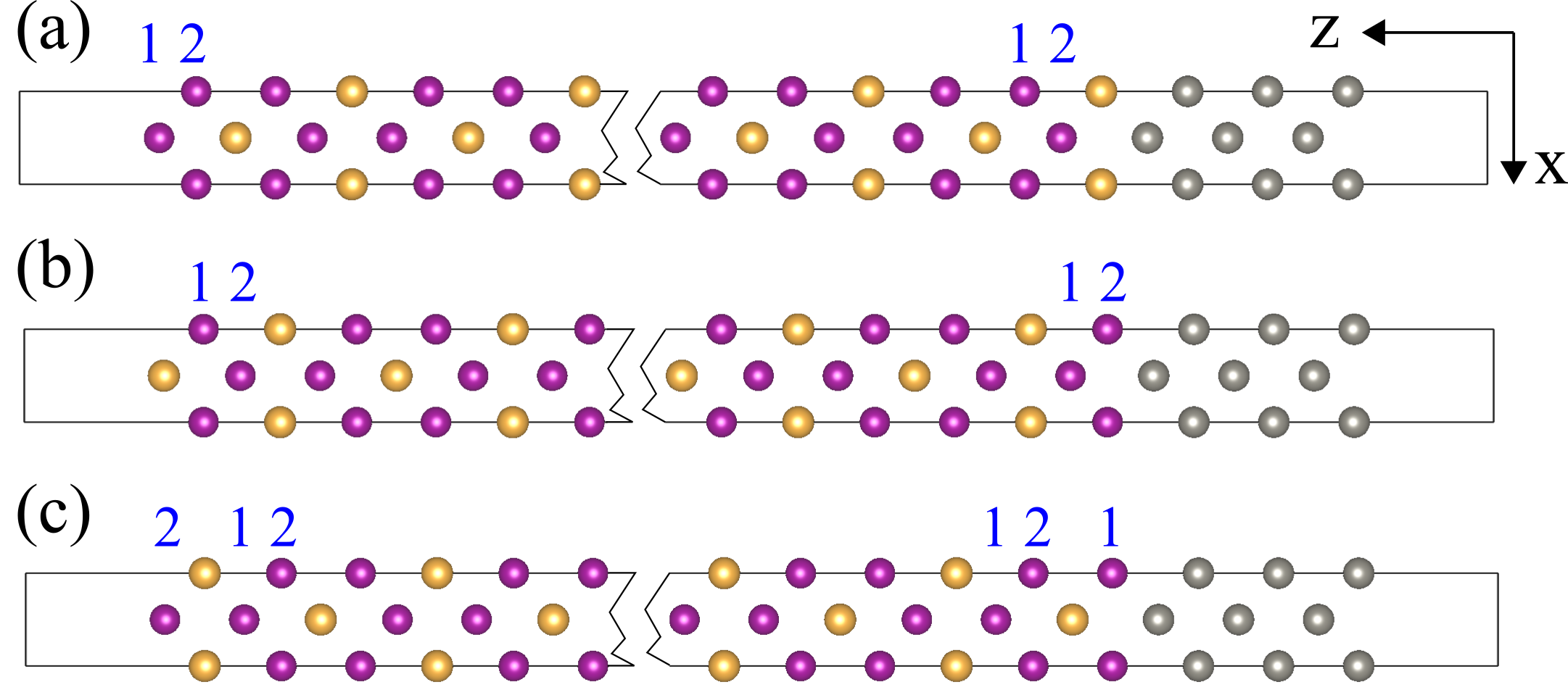}
\caption{Structure of the Mn$_2$Au/W bilayer with (a) termination 1, (b) termination 2, and (c) termination 3. Purple, yellow, and grey balls denote the Mn, Au, and W atoms, respectively. The numbers 1 and 2 label the magnetic sublattices.}
\label{fig:structure}
\end{figure}

We consider a Mn$_2$Au/W bilayer with 54 monolayers (ML) of Mn$_2$Au (36 Mn and 18 Au) and 6 ML of body-centered cubic $\alpha$-W, which we chose because of its simple structure and a relatively small lattice mismatch (about 5\%) with Mn$_2$Au. 
Three types of interface terminations were considered, as shown in Fig. \ref{fig:structure}. Terminations 1, 2, and 3 have, respectively, a monolayer of Au, a double monolayer of Mn, and a single monolayer of Mn as the terminal layer in contact with W. Note that, in order to maintain the stoichiometry of each film, the terminations at the free surface were chosen to be complementary to the termination of the Mn$_2$Au/HM interface. A rather large thickness of Mn$_2$Au was used to separate the effects of the free surface and the interface with HM.

The in-plane lattice constant was fixed at \SI{3.328}{\angstrom}, and the interlayer spacing between Mn$_2$Au and W was relaxed in a unit cell with 2 formula units using the the projector-augmented wave (PAW) method \cite{BLOCHL1994} implemented in the Vienna Ab Initio Simulation Package (VASP). \cite{VASP1,VASP2,VASP3} The interlayer spacings inside Mn$_2$Au and W were taken from bulk relaxed structures with the same constrained in-plane lattice constant. Because fcc Pt structure does not match well with Mn$_2$Au, to address the effect of the HM identity we considered a hypothetical Mn$_2$Au/Pt structure in which W atoms were replaced with Pt while keeping their positions frozen. Given that the true interfacial structure is not known, this will serve as a qualitative comparison of two heavy metals with opposite signs of the spin-Hall effect.

\begin{figure}[htb]
\includegraphics[width=1\columnwidth]{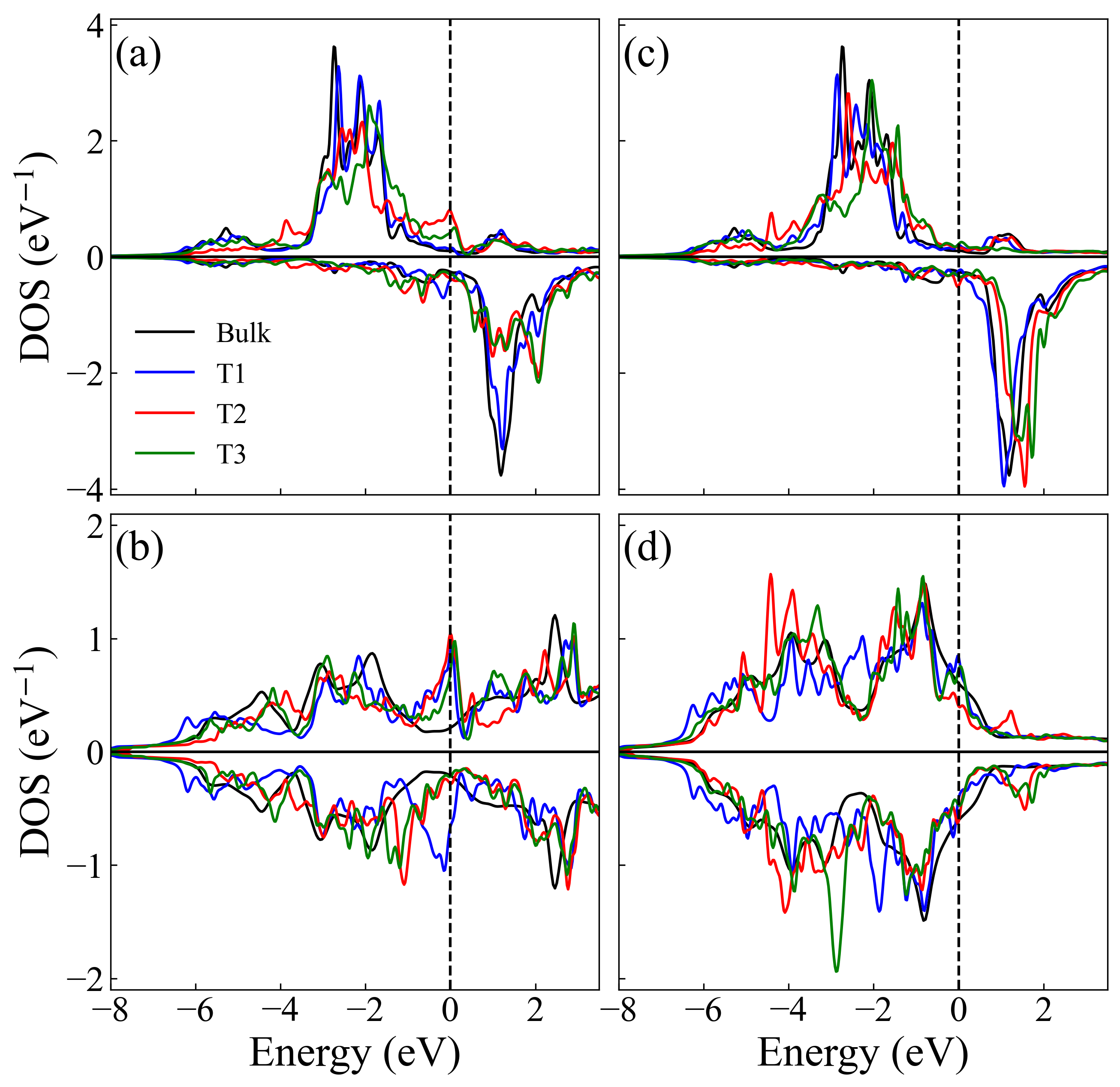}
\caption{Partial density of states of (a) Mn atoms in the Mn$_2$Au/W bilayer, (b) W in Mn$_2$Au/W, (c) Mn in Mn$_2$Au/Pt, and (d) Pt in Mn$_2$Au/Pt. The legend in panel (a) applies to all panels. Blue, red, and green lines: closest atom of the given element to the interface of termination 1, 2, or 3, respectively. Black lines: atom of the given type in bulk Mn$_2$Au, W, or Pt.}
\label{fig:dos}
\end{figure}

The partial densities of states (PDOS) of the Mn, W, and Pt atoms closest to the Mn$_2$Au/HM interface are shown in Fig. \ref{fig:dos} for all three terminations. Because the systems are all-metallic, all PDOS are quite similar to their bulk counterparts. However, Fig. \ref{fig:dos}(b) shows that the interfacial W atom has a peak at the Fermi level in the spin-up channel (i.e., majority spin of the nearest Mn layer) for all three terminations, suggesting the existence of a broadly dispersing interface resonant band. A similar peak is found in the spin-down channel at a slightly lower energy for termination 1, and about 1 eV lower in energy for terminations 2 and 3. Except for the spin-up channel for termination 1, these states hybridize with the nearest Mn atoms, as seen from the presence of similar PDOS peaks in Fig. \ref{fig:dos}(a). Hybridization is considerably suppressed by the intervening Au layer for termination 1, which is clear from the lack of hybridization of the interface states with Mn in the spin-up channel and from the considerably smaller spin splitting of these states in W.
In Mn$_2$Au/Pt the differences in PDOS for different terminations are much smaller, reflecting weaker hybridization between Mn and Pt states.

The magnetic moments of Mn atoms in the bulk are close to 3.5 $\mu_B$; they are slightly smaller at the interface and slightly larger at the free surface. The interfacial W atoms have sizeable spin moments that are antiparallel to the spins of the nearest Mn atoms, amounting to 0.28, 0.29, and 0.45 $\mu_B$ for terminations 1, 2, and 3. In contrast, interfacial Pt atoms have small spin moments of 0.04, 0.02, and 0.04 $\mu_B$ that are parallel to those on the nearest Mn atoms.

For SOT calculations we used the NEGF technique implemented within the tight-binding linear muffin-tin orbital (TB-LMTO) method \cite{AndersenLMTO,Turek} in the Questaal code. \cite{Faleev2005,QUESTAAL,Kirill1,Kirill2} The vacuum region separating the outer surfaces of the bilayer was represented by four layers of empty spheres. The voltage drop was applied along the [100] direction. Disorder was treated explicitly within the Anderson model with a uniformly distributed random potential $V_i$, $-V_m<V_i<V_m$, applied on each atomic site $i$. The length and width of the active region were 120 ML (19.9 nm) and 2 ML (0.32 nm), respectively, and 36 disorder configurations were used for the averaging. We considered five magnitudes of $V_m$: 0.68, 0.75, 0.82, 0.88, and 0.95 eV, which yield the resistivity of 12, 14, 18, 22, and \SI{29}{\micro\ohm\centi\meter}, respectively. Experimentally, resistivity of \SI{20}{\micro\ohm\centi\meter} at 300 K was reported for bulk Mn$_2$Au. \cite{resistivity} The electric field in the embedded region is determined as $E=VGdR/dL$, where $V$ is the voltage drop, $G$ the Landauer-B\"uttiker conductance of the supercell embedded between the two leads, $R=1/G$, and $L$ the length of the active region.\cite{CoPt_2021} 

Site-resolved torquances are defined as $\boldsymbol{\tau}_i(\mathbf{n})=\mathbf{T}_i(\mathbf{n})/E$, where $\mathbf{T}_i(\mathbf{n})$ is the torque on site $i$ for the given orientation of the AFM order parameter $\mathbf{n}=(\mathbf{m}_1-\mathbf{m}_2)/2$, with the antiparallel unit vectors $\mathbf{m}_{1,2}$ representing the sublattice magnetizations. The torquances were calculated for 32 orientations of $\mathbf{n}$ and projected on the orthogonal basis set of real vector spherical harmonics, the first two of which represent DL and FL torque components. \cite{Kirill2} All terms beyond DL and FL were found to be negligible. The resulting torquances correspond to the following form:
\begin{equation}
\boldsymbol{\tau}_i(\mathbf{n})=\tau_{\mathrm{DL},i}\mathbf{n}\times(\mathbf{y}\times\mathbf{n})+\tau_{\mathrm{FL},i}\mathbf{n}\times\mathbf{y}.
\end{equation}

\begin{figure}[htb]
\includegraphics[width=0.9\columnwidth]{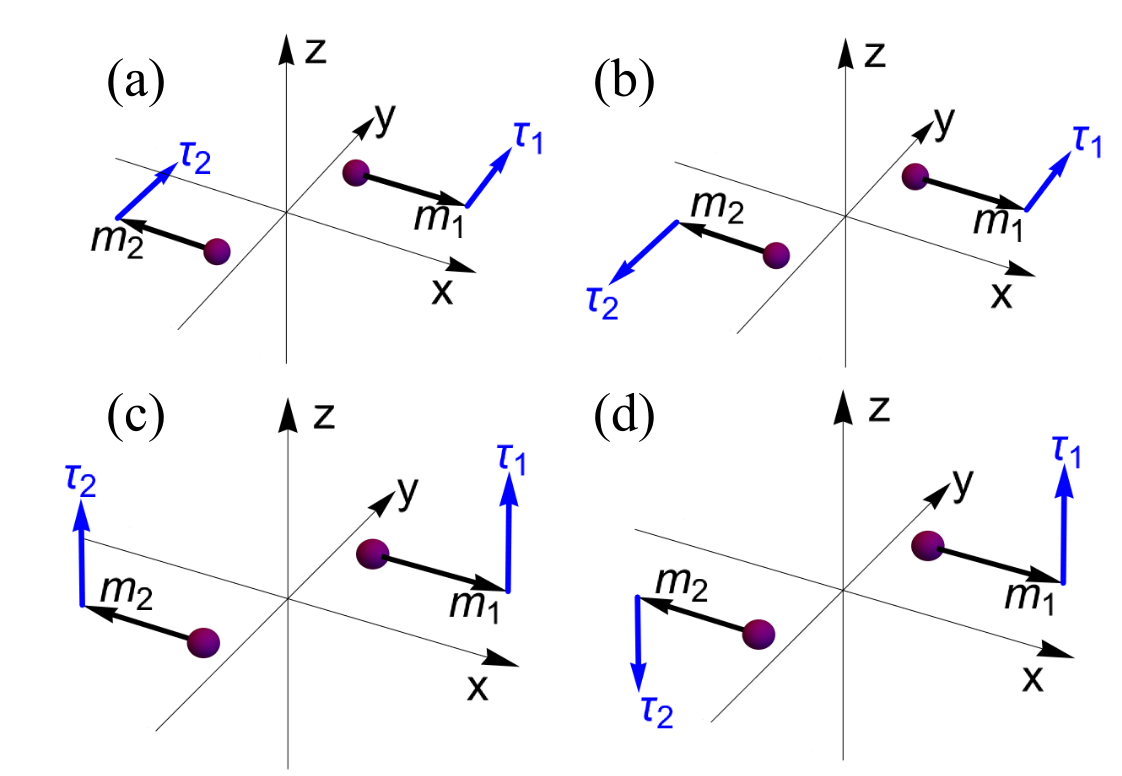}
\caption{Four types of torque illustrated for $\mathbf{n}=\mathbf{x}$: (a) PDL (b) APDL (c) PFL (d) APFL.}
\label{fig:fourtorques}
\end{figure}

The DL and FL torques can be separated into parallel and antiparallel components \cite{Gomonay1,Manchon2017,Baltz-RMP} defined as $\tau_{\mathrm{PDL},k}=(\tau_{\mathrm{DL},k_1}+\tau_{\mathrm{DL},k_2})/2$, $\tau_{\mathrm{APDL},k}=(\tau_{\mathrm{DL},k_1}-\tau_{\mathrm{DL},k_2})/2$, where $k$ labels pairs of adjacent Mn monolayers, and $k_1$, $k_2$ denote the sublattices 1 and 2 within the $k$-th pair, as shown in Fig. \ref{fig:structure}. Similar definitions are used for the PFL and APFL torque components. The four types of torque are illustrated in Fig. \ref{fig:fourtorques} for $\mathbf{n}=\mathbf{x}$. Total torquances for the entire film are also described by efficiencies $\xi^E_{\nu}=(2e/\hbar)\tau_{\nu}/A$ where $A$ is the interfacial area and $\tau_\nu$ is the torquance of type $\nu$ (DL or FL) summed up over all lattice sites.

\section{Results and discussion}

Before discussing the results for specific interface terminations, we remark that the total torques of all four types shown in Fig. \ref{fig:fourtorques} are expected to be finite in a single-crystal Mn$_2$Au/HM bilayer even if it has a rough interface. In other words, the contributions from different surface terminations do not average out to zero. This is because the magnetic space group of Mn$_2$Au does not contain anti-translations (i.e., lattice translations combined with time reversal) which would generate equivalent surface terminations with spin configurations related to each other by time reversal. In an AFM/HM bilayer with an AFM material that does have an anti-translation in the symmetry group (such as MnPt), one can show that the total torques acting on sublattices 1 and 2 should be related as $\boldsymbol{\tau}_1(\mathbf{n})=\boldsymbol{\tau}_2(-\mathbf{n})$ if all equivalent interface terminations are equally represented. This implies that APDL and PFL torques would be zeroed out by interface roughness in a system with such symmetry. We emphasize that this is \emph{not} the case in Mn$_2$Au/HM: even with a rough interface, Mn$_2$Au carries a boundary magnetization \cite{Belashchenko2010} and the associated ferromagnet-like responses.

\begin{figure*}[htb]
\includegraphics[width=0.9\textwidth]{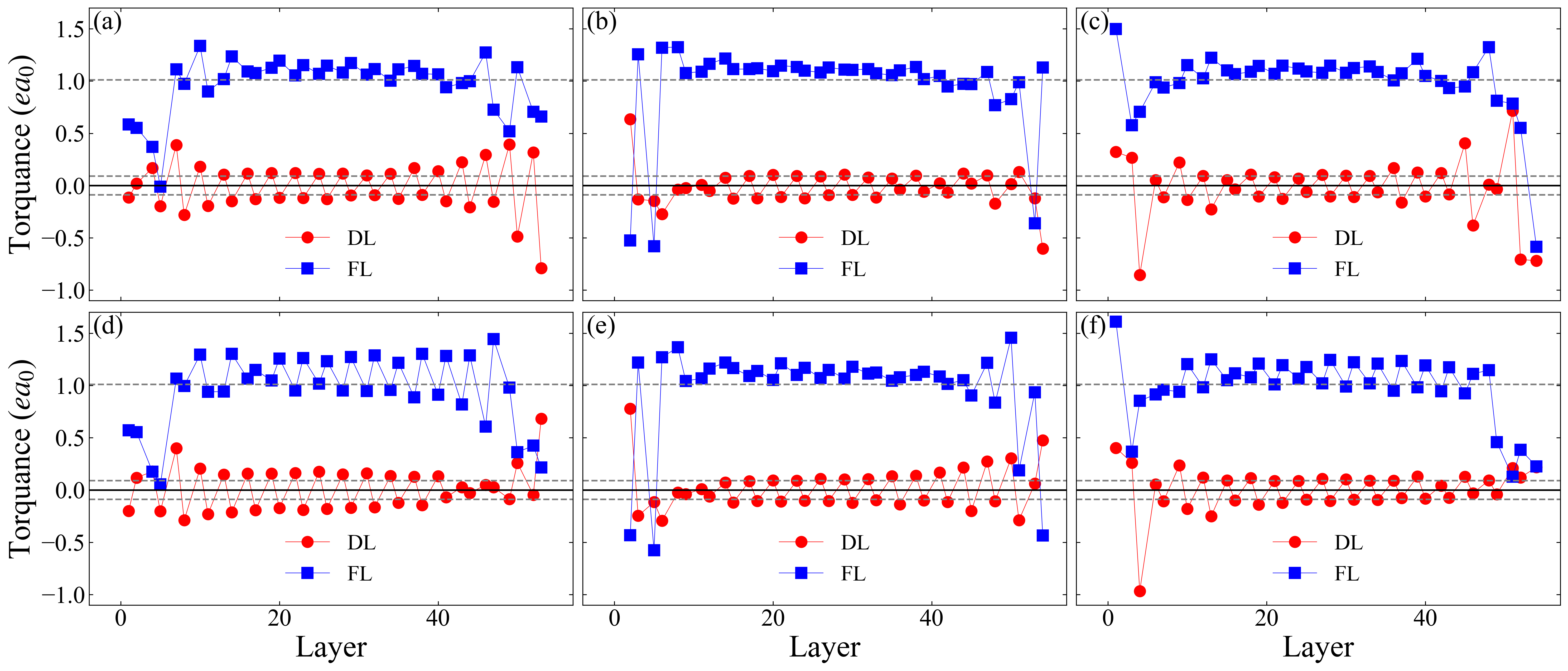}
\caption{Atom-resolved DL (blue squares) and FL (red circles) SOT on the Mn atoms in (a-c) Mn$_2$Au(18 f.u.)/W(6 ML) and (d-f) Mn$_2$Au(18 f.u.)/Pt(6 ML) bilayers with $V_m=0.68$ eV. (a) and (d): termination 1; (b) and (e): termination 2; (c) and (f): termination 3. The $x$ axis shows the layer number in Mn$_2$Au, counting both Mn and Au layers. Layers are counted from the free surface. Dashed lines: PFL and APDL torques from separate bulk calculations for Mn$_2$Au.}
\label{fig:SOT}
\end{figure*}

Figure \ref{fig:SOT} shows site-resolved torquances $\tau_{\mathrm{DL},i}$ and $\tau_{\mathrm{FL},i}$ in Mn$_2$Au/W and Mn$_2$Au/Pt bilayers with three types of interface termination and disorder strength $V_m=0.68$ eV. In the bulk of the film, FL SOT is mostly parallel and slightly exceeds 1 $ea_0$, while DL SOT is mostly antiparallel and has a much smaller magnitude. PFL and APDL SOT components are allowed by symmetry in bulk Mn$_2$Au;  dashed lines in Fig.\ \ref{fig:SOT} show their magnitudes obtained from a separate bulk calculation.

Within a few layers of both faces of each film, the DL and FL SOT deviate strongly from their bulk behavior. The torques near the free surface (left side) are very similar for Mn$_2$Au/W and Mn$_2$Au/Pt bilayers with the same surface termination, confirming that the film is sufficiently thick to decouple the effects of the two faces. However, the torques are quite different for different terminations of each face. They are also very different near the interfaces with W and Pt. FL SOT in Mn$_2$Au/Pt bilayers with terminations 1 and 3 contains a sizeable AP component throughout the thickness of the film. This component is forbidden by symmetry in bulk Mn$_2$Au and must originate at the interfaces. Its existence is consistent with the spin-diffusion theory. \cite{Manchon2017}

\begin{figure*}[htb]
\includegraphics[width=0.9\textwidth]{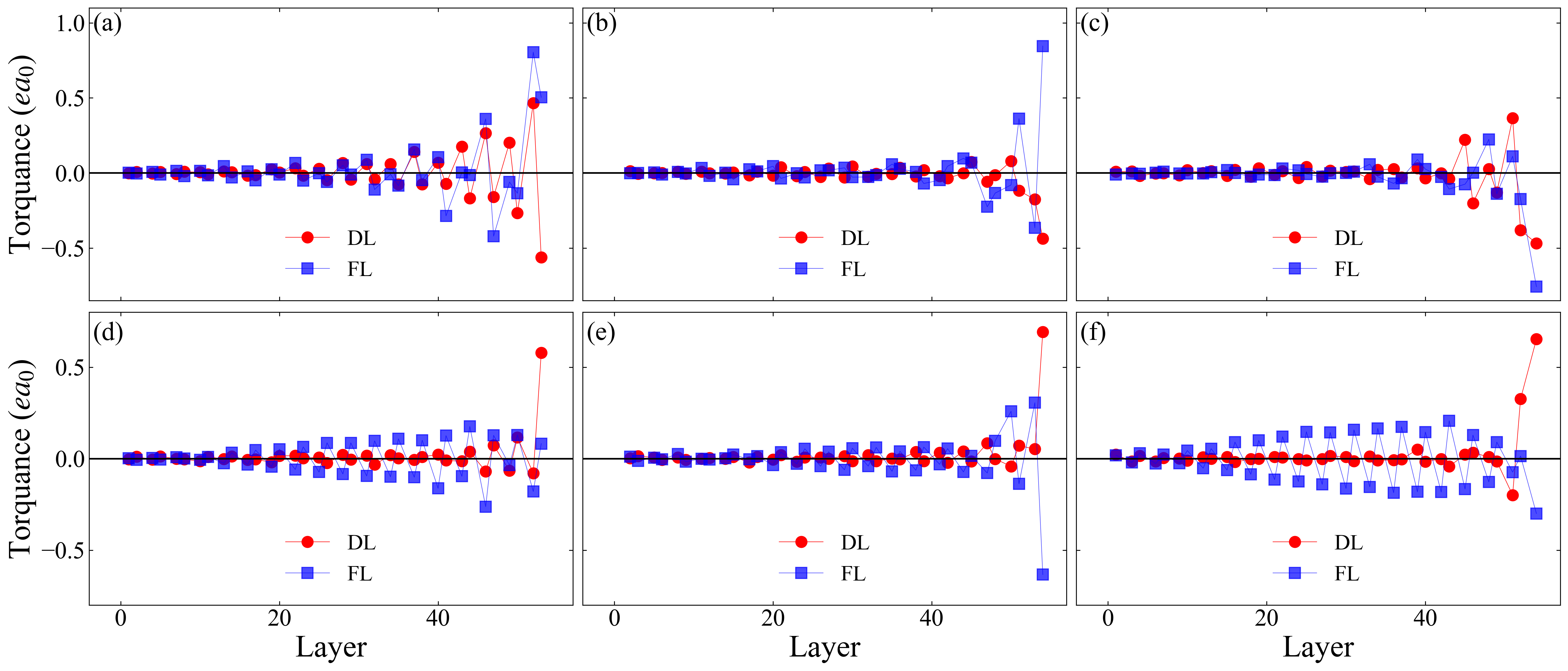}
\caption{Same as in Fig.\ \ref{fig:SOT} but with SOC turned off on all Mn and Au atoms.}
\label{fig:SOT-noSOC}
\end{figure*}

Figure \ref{fig:SOT-noSOC} shows the same quantities as in Fig.\ \ref{fig:SOT} but with SOC turned off on all Mn and Au atoms, which eliminates the bulk contributions to DL and FL SOT. The SOT near the free surface is now close to zero, confirming that the films are thick enough to separate the effects of the two faces. As expected, there are still large DL and FL torques near the interface with W or Pt, which are induced by SOC in the HM layer. The APFL component in Mn$_2$Au/Pt bilayers with terminations 1 and 3 persists without SOC in Mn$_2$Au, showing that it must originate in spin currents coming from the interface with HM and penetrating deep into Mn$_2$Au.

\begin{figure*}[htb]
\includegraphics[width=0.9\textwidth]{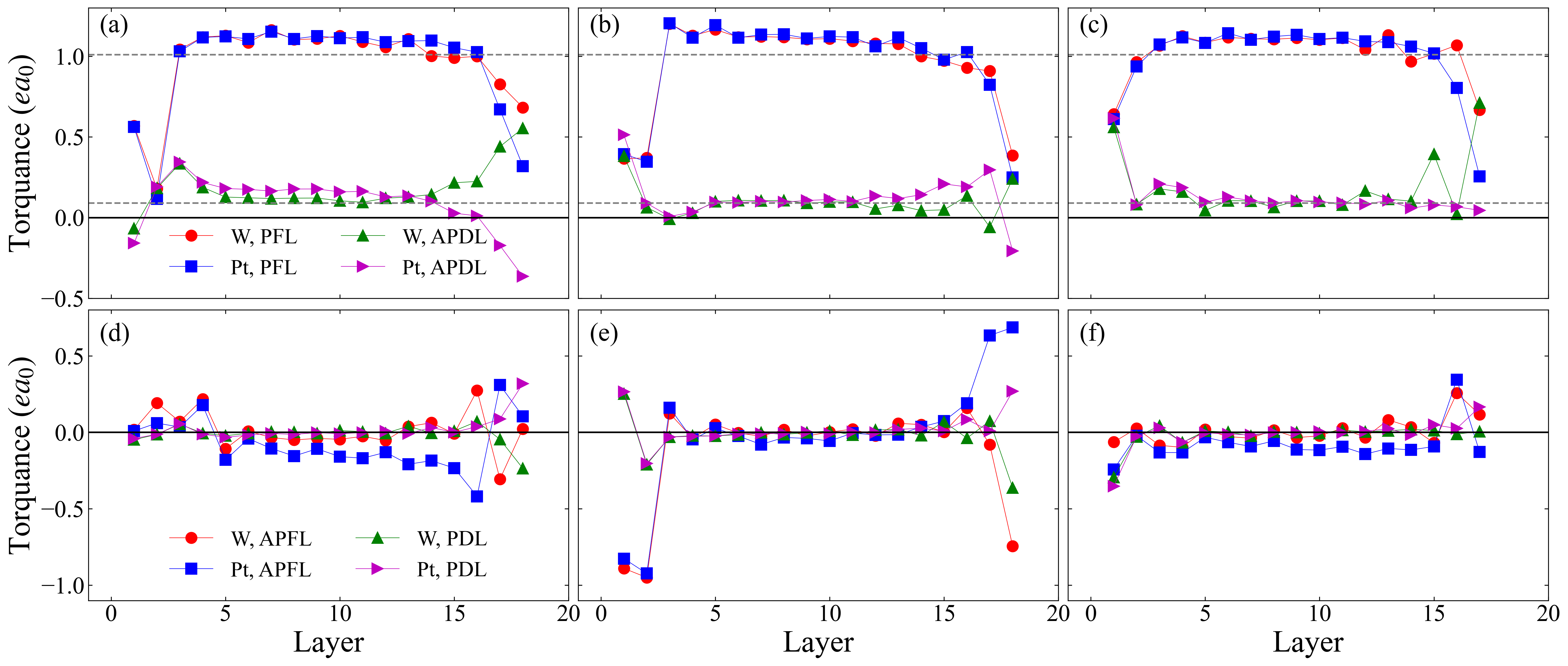}
\caption{Dampinglike and fieldlike torquances in Mn$_2$Au/W and Mn$_2$Au/Pt bilayers decomposed into parallel and antiparallel components. (a) and (d): termination 1; (b) and (e): termination 2; (c) and (f): termination 3. The $x$ axis labels the double layers of Mn. For termination 3, the unpaired Mn layers at each face are omitted. The legend in panel (a) applies to panels (a)-(c), and the legend in panel (d) to (d)-(f). Dashed lines: PFL and APDL torques from separate bulk calculations for Mn$_2$Au.}
\label{fig:sumDiff}
\end{figure*}

\begin{figure*}[htb]
\includegraphics[width=0.9\textwidth]{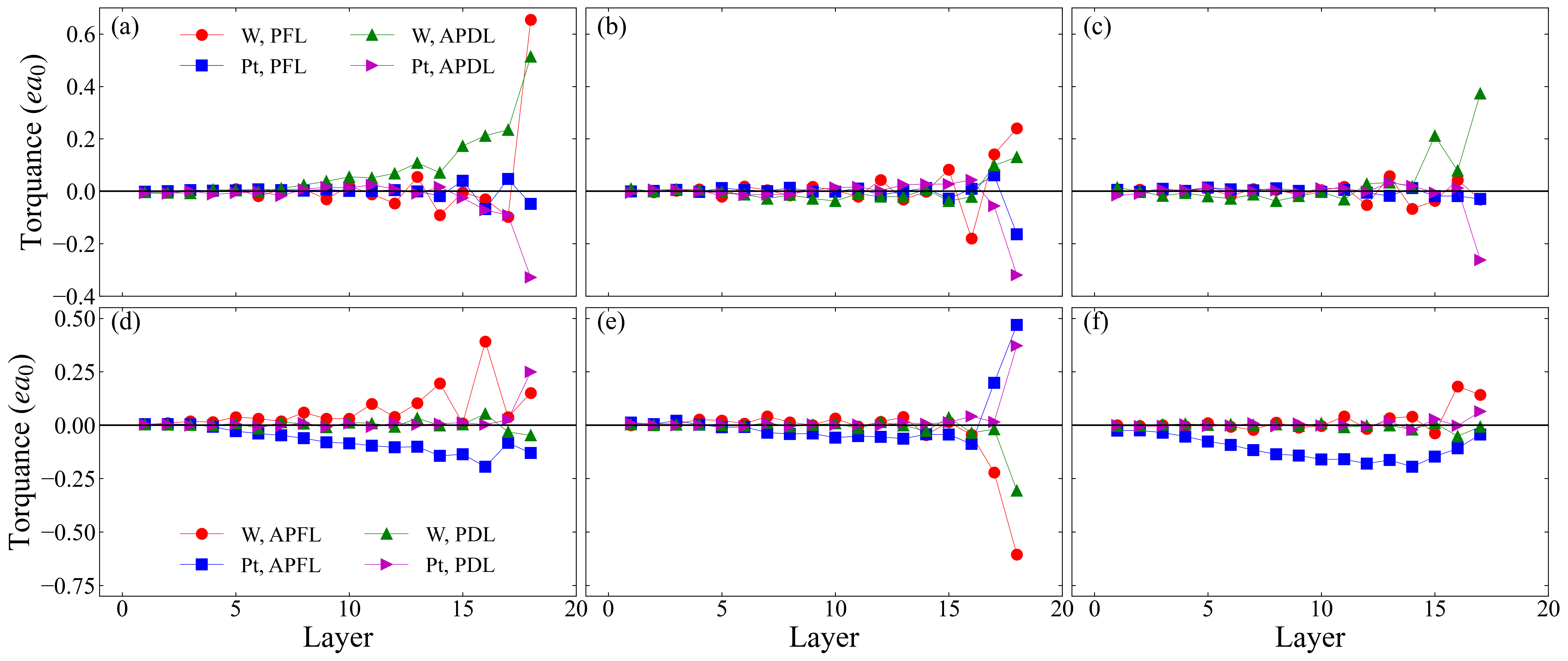}
\caption{Same as in Fig.\ \ref{fig:sumDiff} but with SOC turned off on all Mn and Au atoms.}
\label{fig:sumDiff-noSOC}
\end{figure*}

Figure \ref{fig:sumDiff} exhibits the DL and FL torquances sorted into parallel and antiparallel components, and Fig.\ \ref{fig:sumDiff-noSOC} shows them for bilayers with SOC turned off on all Mn and Au atoms. As seen in Fig.\ \ref{fig:sumDiff} and noted above, the PFL and APDL torques are finite in the bulk of Mn$_2$Au with SOC included. Within a few monolayers of each face the PFL torque is strongly reduced. Interfaces also strongly enhance the APDL and generate PDL and APFL torques with patterns that vary among different terminations. We again note that the APFL torque penetrates deep into Mn$_2$Au in Mn$_2$Au/Pt with terminations 1 and 3. In addition, the APFL torque is considerably enhanced in Mn$_2$Au/W with termination 1 when SOC is turned off in Mn$_2$Au; cf. Fig. \ref{fig:sumDiff}(d) and \ref{fig:sumDiff-noSOC}(d).

Parallel torques are exchange-enhanced \cite{GomonayLoktev2014,Manchon2017,GomonayNatPhys2018} and can efficiently drive magnetization dynamics. Because Mn$_2$Au has a strong bulk PFL torque, it should dominate in thick films. In thin films, however, the PDL torque coming from the interfaces can become comparable to the bulk PFL torque. Surprisingly, it was found using Hall resistance measurements that DL torque from the interface with Pt was strong enough to overcome the bulk FL torque in 10-nm and 25-nm Mn$_2$Au(103) films, resulting in the reorientation of the AFM order parameter along the direction of the current. \cite{Song2019}
Switching by DL torque has also been reported in other metallic antiferromagnets \cite{Dunz2020,Khalili2020,DuttaGupta2020} where symmetry does not allow bulk torques.

The APDL torque was found in first-principles calculations for the antiferromagnetic CrI$_3$ bilayer. \cite{Fei2021} Because the exchange coupling in this system is comparable to magnetocrystalline anisotropy, the APDL torque can affect the dynamics and even switch the AFM order parameter. In Mn$_2$Au, where exchange coupling is much stronger, the APDL torque is not expected to be important for magnetization dynamics.

Figure \ref{fig:SOT} shows that site-resolved DL torquances are large within a few monolayers of each face of the Mn$_2$Au layer, similarly to FM/HM bilayers. \cite{Kirill1} The total DL torque efficiencies $\xi^E_\mathrm{DL}$ are listed in Table \ref{tab:PDL}. We see that $\xi^E_\mathrm{DL}$ is negative for Mn$_2$Au/W and positive for Mn$_2$Au/Pt bilayers with all three terminations, which is consistent with the opposite signs of $\xi^E_\mathrm{DL}$ observed in FM/HM bilayers with both $\alpha$ and $\beta$ phases of W compared to Pt. \cite{PtSHA,ManchonRMP} The magnitude of $\xi^E_\mathrm{DL}$ is similar to FM/HM bilayers. \cite{ManchonRMP,zhu2021maximizing} However, the differences among the three terminations are very large. For Mn$_2$Au/W, the magnitude of $\xi^E_\mathrm{DL}$ for termination 1 is much smaller compared to terminations 2 and 3, whereas for Mn$_2$Au/Pt it is much smaller for termination 3 compared to 1 and 2. 

\begin{table}[htb]
    \centering
    \begin{tabular}{l|c|c|c|c|c|c}
    \hline    
    Bilayer        & \multicolumn{3}{c|}{Mn$_2$Au/W}  & \multicolumn{3}{c}{Mn$_2$Au/Pt}  \\
    \hline
    Termination    & 1      & 2     & 3     & 1     & 2     & 3\\
    \hline
    Total & -0.95 & -1.51 & -2.37 &  1.74 & 1.69  &  0.55\\
    \hline 
    (L) Free surface & -0.04 & -0.03 & -0.71 & -0.04 & 0.04  & -0.84\\
    \hline
    (R) Interface with HM & -1.02 & -1.31 & -1.65 &  2.05 & 1.73  &  1.54\\
    \hline
    Sum of L and R   & -1.06 & -1.34 & -2.36 &  2.01 & 1.77  &  0.70\\
    \hline
    SOC only in HM & 0.13 & -1.56 & -1.39 & 1.44 & 2.21 & 1.92\\
    \hline
    SOC only in Mn$_2$Au & -0.34 & -0.57 & -0.73 & -0.28 & -0.83 & -0.35\\
    \hline
    \end{tabular}
    \caption{DL torque efficiency $\xi^E_\mathrm{DL}$ (units of \SI{e5}{\per\ohm\per\meter}) for $V_m=0.68$ eV and its decomposition into contributions from the free surface and the interface with HM (see text). The last two lines list $\xi^E_\mathrm{DL}$ obtained with SOC turned on only in HM or only in Mn$_2$Au.}
    \label{tab:PDL}
\end{table}

Because there is a finite APDL torque in the bulk of Mn$_2$Au, the separation of the total DL torquance $\xi^E_\mathrm{DL}$ into contributions from the two faces is not unique. We employed a summation regularized by weighting with the smeared step function $F(n_i)=\{1+\exp[(n_i-d_0)/\Delta]\}^{-1}$, where $n_i$ is the layer index measured from the given face, and we chose $d_0=15$ and $\Delta=2$. Table \ref{tab:PDL} shows the contributions from the left and right faces obtained in this way along with their sum. In all cases this sum is close to the total DL torque efficiency, as expected from the near-vanishing of the PDL torquance in the middle of each device (as seen in Fig.\ \ref{fig:sumDiff}).

The results of this separation suggest that there is a large negative contribution to $\xi^E_\mathrm{DL}$ from the free surface with termination 3. Large DL SOT at the free surface was also found in calculations for Co; \cite{Kirill1} it is related to the anomalous SOT observed experimentally at the ferromagnetic surfaces and could be due to the spin-Hall effect inside the magnetic layer. \cite{Wang2019} Strong dependence of the free-surface DL SOT on the termination suggests that its origin in Mn$_2$Au is more complicated. For example, the conversion of the spin current into torque can be strongly modified by SOC at the surface.

The contribution from the interface with the HM exhibits a large variation for different terminations.
Some of this variation can be due to the differences in the spin-mixing conductance and spin memory loss at the interface, \cite{zhu2021maximizing} but
it can also be due to other mechanisms beyond the spin-Hall effect in HM contributing to SOT. Such mechanisms may include spin currents generated at the interfaces, \cite{Amin.Stiles.PRB2016,Amin.Stiles.PRB2016a} spin-Hall current generated in the magnetic material \cite{Amin2019} and absorbed at the interfaces or in the HM, or orbital Hall current \cite{Tanaka-OHE,Kontani-OHE,Go-OHE} from the normal metal converted by SOC into SOT. The contributions from these mechanisms depend on the properties of the interfaces. \cite{Lee2021,go2021longrange}
First-principles calculations showed a giant enhancement of the spin-Hall angle near the Co/Pt interface, \cite{Kelly2016} and a large interfacial contribution to DL torquance was identified for ferromagnetic Co/Pt and Co/Au bilayers through the analysis of its thickness dependence. \cite{Kirill2}

In Mn$_2$Au, SOC effects are expected to be stronger compared to $3d$ metals, and evidence of spin-Hall effect in Mn$_2$Au was recently reported. \cite{Singh2020,Chen2021}
Table \ref{tab:PDL} lists the DL efficiencies $\xi^E_\mathrm{DL}$ obtained with SOC turned on only in the HM layer or only in Mn$_2$Au. By comparing the sum of these two results with $\xi^E_\mathrm{DL}$ obtained with SOC included everywhere, we see that the effects of spin-orbit coupling in the two layers are not additive. Spin memory loss at the interface could lead to the non-additivity of SOC effects, but it can not explain why in some cases (Mn$_2$Au/W with terminations 1 and 3; Mn$_2$Au/Pt with termination 1) the torque efficiency $\xi^E_\mathrm{DL}$ is enhanced by SOC in Mn$_2$Au, or why it changes sign in Mn$_2$Au/W with termination 1. The sizable $\xi^E_\mathrm{DL}$ found with SOC turned on only in Mn$_2$Au (i.e., the self-torque, \cite{Wang2019} which here can be due in part to the orbital Hall effect) also shows that the DL torque appears in part due to mechanisms that are unrelated to the spin-Hall effect in HM.

\begin{figure}[htb]
\includegraphics[width=0.9\columnwidth]{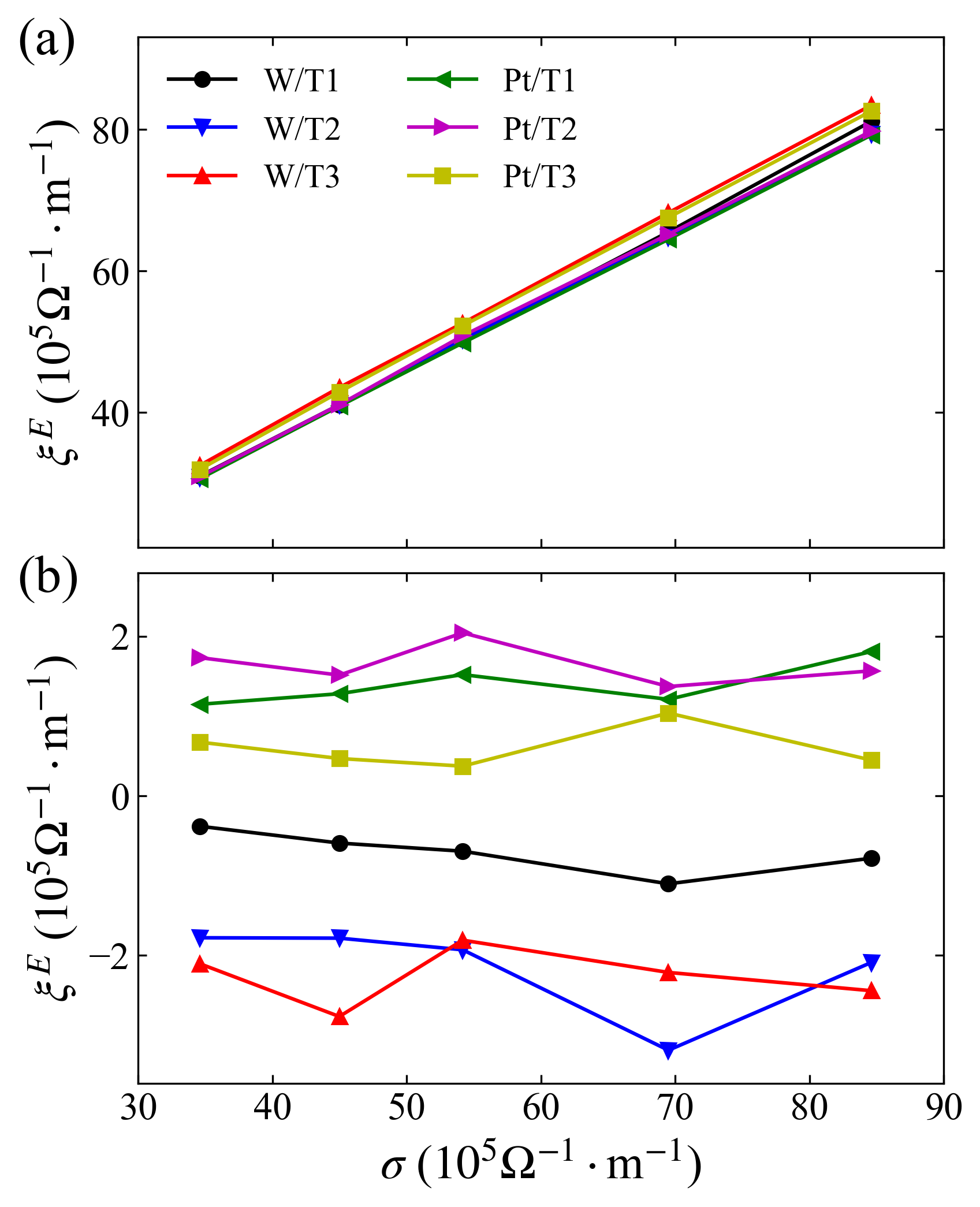}
\caption{(a) $\xi^E_\mathrm{FL}$ and (b) $\xi^E_\mathrm{DL}$ torque efficiencies in Mn$_2$Au/W and Mn$_2$Au/Pt as a function of the effective conductivity. The legend lists the heavy metal and the termination type.}
\label{fig:disorderNS}
\end{figure}

Figure \ref{fig:disorderNS} shows the dependence of $\xi^E_\mathrm{FL}$ and $\xi^E_\mathrm{DL}$ on the conductivity, which is controlled by the disorder strength $V_m$. The results presented above correspond to the largest conductivity shown in these figures. The FL efficiency $\xi^E_\mathrm{FL}$ in our thick film is dominated by bulk torque, and it scales linearly with the conductivity, which is consistent with the inverse spin-galvanic mechanism. In contrast, $\xi^E_\mathrm{DL}$ does not exhibit a clear trend as a function of $\sigma$, suggesting that DL torque is dominated by mechanisms that are insensitive to disorder strength. 

\begin{figure}[htb]
\includegraphics[width=0.9\columnwidth]{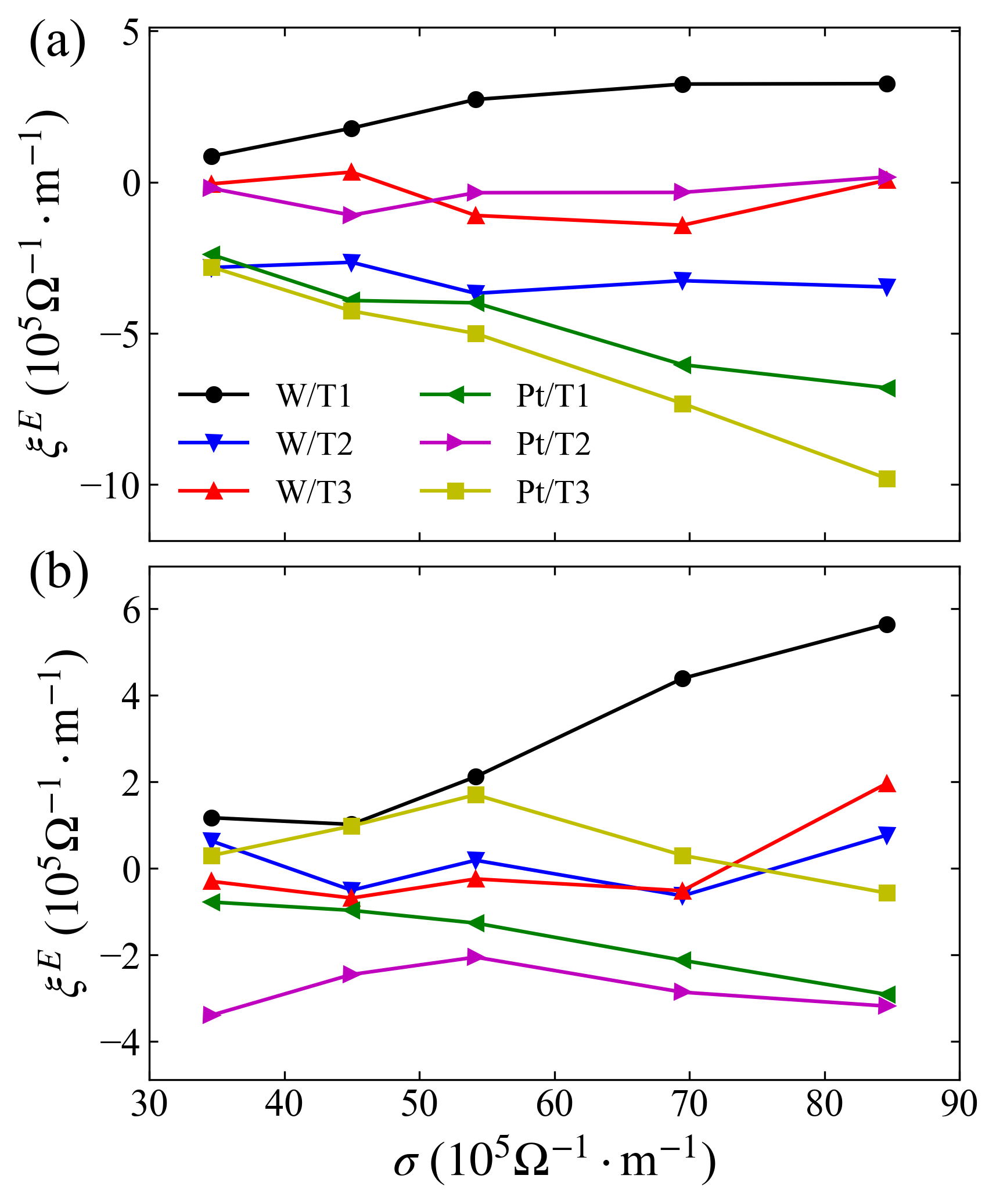}
\caption{(a) APFL SOT and (b) APDL SOT in Mn$_2$Au/W and Mn$_2$Au/Pt with SOC turned off on all Mn and Au atoms, as a function of the effective conductivity. The legend lists the heavy metal and the termination type.}
\label{fig:disorderS}
\end{figure}

Figure \ref{fig:disorderS} shows the effect of disorder strength on APFL and APDL torque efficiencies with SOC turned off on all Mn and Au atoms. Antiparallel torque efficiencies are defined in the same way as conventional $\xi^E_\mathrm{DL}$ and $\xi^E_\mathrm{FL}$ but with the site-resolved torques on the two sublattices added with opposite signs. The SOC has been turned off in Mn$_2$Au in order to exclude the bulk APDL torque and focus on interface-generated torques. As we noted above in Fig.\ \ref{fig:SOT}-\ref{fig:sumDiff-noSOC}, there is a fairly strong APFL SOT in Mn$_2$Au/Pt bilayers with terminations 1 and 3, which is attributable to the nonstaggered spin accumulation predicted by the spin-diffusion theory. \cite{Manchon2017} There is also a relatively large APFL torque in Mn$_2$Au/W with termination 1 if SOC is turned off in Mn$_2$Au, although its decay into the depth of Mn$_2$Au is not as smooth [see Fig. \ref{fig:sumDiff-noSOC}(d)]. In these three cases, we see from Fig.\ \ref{fig:disorderS}(a) that APFL SOT increases with increasing conductivity $\sigma$, which can be due to the increasing transverse spin-diffusion length \cite{Manchon2017} in Mn$_2$Au. In other cases the APFL torque is sizeable only near the interface (see Fig.\ \ref{fig:sumDiff-noSOC}) and does not exhibit a clear trend as a function of disorder strength. In its effect on spin dynamics, APFL torque is equivalent to a magnetic field applied along the $y$ axis.

As seen in Fig.\ \ref{fig:SOT}-\ref{fig:sumDiff-noSOC}, some interfaces generate strong APDL torque, which is not present in the spin-diffusion theory of Ref.\ \onlinecite{Manchon2017}. This component is present in Mn$_2$Au because the equivalence of the two AFM sublattices is broken near any (001) interface. This sublattice symmetry breaking is required by bulk symmetry even in the presence of interface roughness in thermodynamic equilibrium. \cite{Belashchenko2010} As a result, the spin current incident from the HM interacts unequally with the two sublattices, and the transferred angular momentum contributes to both PDL and APDL torques. In order to include this contribution to APDL, the spin-diffusion theory \cite{Manchon2017} would need to include a sublattice-asymmetric boundary condition at the AFM/HM interface.

Inspection of Fig.\ \ref{fig:SOT-noSOC} shows that strong interface-generated DL torque is usually seen only on 1-3 layers of Mn near the interface. One such example is the Mn$_2$Au/Pt interface with termination 2, where the DL torque is large on one Mn layer near the surface and small everywhere else. Figure \ref{fig:disorderS}(b) shows that the total APDL torque for this interface does not exhibit a clear trend as a function of conductivity, similarly to the PDL torque in Fig. \ref{fig:disorderNS}(b). This similarity reflects the fact that PDL and APDL torques in this case are dominated by the same layer of Mn atoms. In contrast, in Mn$_2$Au/W with termination 1 there is a relatively strong oscillating DL torque that penetrates several unit cells deep into Mn$_2$Au. A similar but less pronounced feature is seen in Mn$_2$Au/Pt with the same termination. In both these cases, Fig.\ \ref{fig:disorderS}(b) shows that the APDL torque is increasing with increasing conductivity. This trend is similar to the APFL torque for Mn$_2$Au/Pt bilayers with terminations 1 and 3, and it can also be explained by the disorder dependence of the spin-diffusion length.

\section{Conclusions}

We have studied SOT in Mn$_2$Au/HM bilayers (where HM is W or Pt) using the NEGF approach with explicit averaging over Anderson disorder. The SOT is dominated by the well-known bulk fieldlike torque that is parallel on the two sublattices and scales linearly with the conductivity. There is also a weaker bulk dampinglike torque that is antiparallel on the two sublattices. The interface with a heavy metal generates disorder-insensitive parallel dampinglike torque with an efficiency $\xi^E_\mathrm{DL}$ that is comparable to ferromagnetic bilayers.
This efficiency strongly depends on the interface termination and sometimes changes drastically if SOC is turned off in Mn$_2$Au, showing that the DL torque is not due solely to the spin-Hall current generated in the HM layer. Some interfaces induce sizeable antiparallel fieldlike torque which penetrates deep into Mn$_2$Au, in agreement with the spin-diffusion model. These results can help in the design and optimization of antiferromagnetic spintronic devices.

\begin{acknowledgments}
We are grateful to Aur\'elien Manchon and Xin Fan for useful discussions. This work was supported by the National Science Foundation through Grant No. DMR-1916275. Calculations were performed utilizing the Holland Computing Center of the University of Nebraska, which receives support from the Nebraska Research Initiative.
\end{acknowledgments}

\bibliography{SOT}

\end{document}